%Paper: cond-mat/9301032
%From: leibig@hlrserv.hlrz.kfa-juelich.de (Michael Leibig)
%Date: Tue, 26 Jan 93 12:02:50 +0100
%Date (revised): Tue, 26 Jan 93 23:42:11 +0100

% To Tex: requires new harvard macros (9/91 update)
\input harvmac
%\abovedisplayskip = 0 true cm
%\abovedisplayshortskip = 0 true cm
\lref\prlliu{C.\ Liu and S.\ Nagel,  Phys.\ Rev.\ Lett.\
{\bf 86}, 2301 (1992).}

\lref\landlif{L.\ Landau and E.\ Lifshitz, {\it Theory of Elasticity},
Pergammon Press, London, 1959.}

\lref\review{H. Jaeger and S. Nagel, Science {\bf 255}, 1523 (1992)}.

\lref\bcgref{W.\ Press and S.\ Teukolsky, Computers in Physics {\bf
6}, 400 (1992).}

\lref\numrec{W.\ Press, B.\ Flannery, S. Teukolsky and W. Vetterling,
{\it Numerical Recipes: The Art of Scientific Computing}, Cambridge
U.\ P., New York, 1988.}

\lref\phonref{A. Czachor, {\it Physics of Phonons: Proceedings of the
XXIII Winter School of Theoretical Physics\/}, ed.\ T.\ Paszkiewicz,
Springer-Verlag, Berlin (1987).}

\lref\convref{P. Evesque and J.\ Rajchenbach, Phys.\ Rev.\ Lett.\ {\bf
62}, 44 (1989).}

\lref\avalref{P.\ Bak, C.\ Tang and K.\ Weisenfeld, Phys.\ Rev.\
Lett.\ {\bf 59}, 381 (1987).}

\lref\archref{M.\ Ammi, D.\ Bideau and J.\ Troadec, J.\ Phys.\ D {\bf
20}, 424 (1987).}

\lref\wavesref{G.\ Baxter, R.\ Behringer, T. Fagert and G. Johnson,
Phys.\ Rev.\ Lett.\ {\bf 62}, 2825 (1989).}

\lref\localref{B.\ Souillard in {\it Chance and Matter, Les Houches 1986,
Session XLVI\/}, eds.\ J.\ Souletie, J.\ Vannimenus and R.\ Stora, North
Holland, Amsterdam, 1987, and references therein.}

\lref\moderev{P. Dean, Rev.\ Mod.\ Phys.\ {\bf 44}, 127 (1972), and
references therein.}

\lref\modjones{S.\ Nagel, S.\ Rahman and G.\ Grest, Phys.\ Rev.\
Lett.\ {\bf 47}, 1665 (1981)}.

\lref\modperc{A.\ Petri and L.\ Pietronero Phys.\ Rev.\ B {\bf 45},
12864 (1992), and references therein.}

\lref\localdrive{L.\ Ye, G.\ Cody, M.\ Zhou, P.\ Sheng and A.\ Norris,
Phys.\ Rev.\ Lett.\ {\bf 69}, 3080 (1992).}

\lref\linsysref{J. Marion, {\it Classical Dynamics of Particles and
Systems}, Academic Press, Orlando, 1977.}

\lref\liuunp{C.\ Liu and S.\ Nagel, unpublished.}

\lref\nagelsand{H.\ Jaeger, C. Liu and S.\ Nagel, Phys.\ Rev.\ Lett.\ {\bf
62}, 40 (1989)}

\lref\andloc{P.\ Anderson, Phys.\ Rev.\ {\bf 109}, 1492 (1958).}

\def\delh{\delta h}
\def\delF{\delta F}
\def\eqK{K_h}
\def\Kmx{K_{max}}
\def\omx{\omega_{max}}
\def\eqKpp{K_h(p, p')}
\def\Kz{K_0}
\def\up{u(p)}
\def\upp{u(p')}
\def\pp{p'}
\def\Ap{A(p)}
\def\App{A(p')}
\def\Axy{A(x,y)}
\def\vph{v_{\phi}}
\def\slope{\mu}

\def\ni{\noindent}
\def\aln{\alpha_n}
\def\xy{(x,y)}
\def\ompk{\omega_{p}}

\def\fixy{f_i(x,y)}
\def\omi{\omega_i}
\def\eigfun{f_i(x,y)}

\def\powspc{\psi}

\hsize=15.5truecm
\hoffset=0 true cm     % This is set to compensate for the HLRZ printer
\voffset=0 true cm     % This is set to compensate for the HLRZ printer
\vsize=22.5truecm

%\draftmode
\Title{HLRZ Preprint 6/93}{\vbox{\centerline{A Model for the Propagation
of}\vskip 12 true pt
\centerline{Sound in Granular Materials}}}
\baselineskip 15 true pt plus 2pt minus 2pt
\centerline{\sl Michael Leibig}
\centerline{\sl H\"ochstleistungsrechenzentrum, KFA}
\centerline{\sl Postfach 1913}
\centerline{\sl 5170 J\"ulich}
\centerline{\sl Germany}
\bigskip
\baselineskip=20pt plus 2pt minus 2pt
\leftline{\bf Abstract}

This paper presents a simple ball-and-spring model for the propagation
of small amplitude vibrations in a granular material. In this model,
the positional disorder in the sample is ignored and the particles are
placed on the vertices of a square lattice. The inter-particle forces
are modeled as linear springs, with the only disorder in the system
coming from a random distribution of spring constants.  Despite its
apparent simplicity, this model is able to reproduce the complex
frequency response seen in measurements of sound propagation in a
granular system. In order to understand this behavior, the role of the
resonance modes of the system is investigated. Finally, this simple
model is generalized to include relaxation behavior in the force
network -- a behavior which is also seen in real granular materials.
This model gives quantitative agreement with experimental
observations of relaxation.

\vfil
\leftline{PACS numbers: 63.50.+x, 05.40.+j, 46.10.+z, 43.40.+s\hfill Jan. 25,
1993}
\eject
\pageno=2
\headline{\tenrm\hfil\folio\hfil}
\newsec{Introduction}

Understanding the properties of granular materials present a difficult
challenge to condensed matter science \review.  A fundamental property
of any material is its response to small amplitude vibrations. For the
case of solids, the behavior of the system is characterized by the
phonon spectrum. These elementary excitations for perfect crystals are
well understood, but for disordered systems, the situation is not so
clear. When disorder is present, localization of these excitation can
occur \andloc, and while much progress has been made in understanding
the phenomena of localization, there are still many fundamental
properties that remain a mystery \localref.

While there have been many studies of the localized modes in
disordered systems \refs{\moderev\modjones{--}\modperc}, there has
been less attention paid to the response of these systems to a
periodic driving force \localdrive.  Generally speaking, for a uniform
periodically-driven linear system in the long time limit, the
frequency of oscillation is that of the forcing \linsysref.  As the
frequency of the driving goes through a resonant frequency, the
amplitude of the motion is increased, and the oscillations takes on
the character of the corresponding normal mode.  Does this picture
remain for the case of a disordered granular material which has
complicated non-linear inter-particle interactions?  In these systems,
there is known to be a network of contact forces with a complex
structure. The phenomena of arching suggest that this system is
supported by a small fraction of grains which have a high
concentration of stress, while most of the particles remain in loose
contact \archref. The response of this system to driving may then
depend wholly on the structure of this network where the stress is
highly concentrated.

In this paper, I discuss a model system of balls and springs that can
reproduce many of the vibrational properties of a real granular system.
The numerical results that come from this model will be compared with
the results from a beautiful set of experiments performed by Liu and
Nagel in which they measured the response of a single ``grain of
sand'' to an external driving force \prlliu. I describe briefly these
experiments and summarize the important results. The experimental
apparatus consisted of a rigid box filled with glass bead of diameter
$d = .5$ cm. The box itself was $28 {\rm\ cm} \times 28 {\rm\ cm}$ and
was filled with beads up to heights of 8 to 15 cm.  The upper layer of
the bead pack was a free surface. At one end of the box was placed an
aluminum disk that was attached to an external speaker by a rigid rod.
The speaker could be driven with varying frequencies and amplitudes.
Inside the bead pack was an accelerometer which was roughly the size
of one bead. This device was sensitive only to horizontal
accelerations, and was unaffected by sound waves propagating through
the surrounding air.  Since this detector was comparable to the size
of a bead, it is effectively measuring the motion of a single particle
under the action of the driving vibration.  The motion of the aluminum
disk was also monitored with an accelerometer attached directly to it.

The disk was driven with an acceleration of the form $A_s \sin(\omega
t)$ and the detector was found to oscillate as $A_d(t) \sin(\omega t +
\phi(t))$, where $A_d(t)$ and $\phi(t)$ were the detector amplitude
and phase shift, respectively. Both of these functions were found to
vary slowly with time. \nfig\exptime{Experimental results for the
amplitude of a bead at a single driving frequency: (a) shows the
amplitude (in units of $g$) as a function of time, (b) is the power
spectra for this time trace. The bead is a distance of 4 cm from the
driving plate. The dotted line has slope of 2. (Data reproduced with
permission of the authors of ref.\ \xref\prlliu.)} Figure
\xfig\exptime(a) shows $A_d(t)$ with a driving frequency $\omega = 4$
kHz, and $A_s = 1.4 g$ where $g$ is the acceleration of gravity.  Note
that the time scale for the changes in $A_d(t)$ is much longer than
the time scale for the oscillation itself.  The power spectrum (shown
in \exptime(b)) of this time series shows a power law region with
exponent $\approx 2$ at frequencies from $10^{-5}$ Hz to $10^{-1}$ Hz.

On time scales of the order of a few oscillations, however, the
amplitude at a given driving frequency is fixed. This inspired a
second set of measurements using the same experimental set-up. With a
fixed acceleration for the driving force, the amplitude of the
detector is measured for various value of $\omega$. Let the response
function, $\eta(\omega) = A_d / A_s$ for a driving frequency,
$\omega$. This ratio is plotted as a function of $\omega$ in
\fig\expfreq{Experimental results for the
frequency response of a bead as a function of driving
frequency. The bead is again a distance of 4 cm from the driving
plate. The two curves result from successive measurements on an
undisturbed system. The second curve is displaced downward so that the
two are distinguishable. (Data reproduced with
permission of the authors of ref.\ \xref\prlliu.)}.
The two curves shown were done in consecutive sweeps through the
frequency range with the measurements being separated by a short
interval in time. The second curve is displaced vertically downward so
that the two curves can be distinguished. If the sample was disturbed
between the measurements, the response curve looked very different for
the second scan of the frequency range. Thus, while the data looks
``noisy,'' the response is actually characteristic of the particular
contact network of the sample.  Thus, \expfreq\ gives
information about a static force network in the bead pack.  Figure
\xfig\exptime, on the other hand, results from the time evolution of
this network.

In the rest of this work, I present the details of a ball and spring
model for the bead pack and present numerical results for a direct
comparison between the model system and the experimental results. I
also study the role that the system's normal modes play in the
frequency response. The rest of the paper will be organized as
follows.  Section 2 contains a discussion and motivation for a ball
and spring model for this system. In section 3, I will consider the
continuum limit of this model and examine the solution to the
differential equations which describe a homogeneous system. Section 4
presents the numerical solution of the discreet model, and compares
these results with the frequency response seen in the experiment.  In
section 5, I present a toy model for the slow amplitude modulations
which are a result of the relaxation of the force network. Finally, in
section 6, I summarize and discuss extensions to this simple model
which must be made to capture more of the physical characteristics of
the real system.

\bigskip

\newsec{Ball and Spring Model}

Consider the bead pack at equilibrium. In this experiment, the only
important forces acting on each bead are the contact forces between
adjacent particles and gravity. The collection of contact forces among
neighboring beads will be referred to as the contact force network, and
I sometimes refer to the force between two connected beads as a
``bond.''

The fundamental unit of this network is the bond between two spheres.
If two spherical beads are compressed together with a contact force
$F_c$, then under the action of this force the distance between the
centers of the two spheres is decreased by an amount $h$. For the case
of linear elasticity and perfectly smooth spheres, these two
quantities are related by
\eqn\hertz{
F_c = k h^{3/2},
}

\ni where $k$ is a proportionality constant depending on the
radii of the two spheres and their material properties \landlif. For
the case of an almost mono-disperse collection of beads, this constant
$k$ can be treated as the same for all contacts.

Consider now an additional force, $\delF$, acting on these two balls.
The spheres will move together by an additional amount $\delh$.  Under
this new force, equation \hertz\ becomes
\eqn\modhertz{
F_c + \delF = k (h + \delh)^{3/2}.
}

\ni For the case of small $\delh$, I can expand the
expression in parenthesis and making use of Eq.\ \hertz, I find
\eqn\basespr{
\delF = {3k h^{1/2} \over 2} \delh \equiv \eqK \delh,}

\ni where $\eqK$ is a constant which depends on the equilibrium
stress on each bond in the network. Equation \basespr\ shows that for
a sufficiently small displacement, $\delh$, the bond between a pair of
connected beads acts like a simple spring.

Unfortunately, \basespr\ is not true for all bonds. There may exist
some beads which in equilibrium are very close to each other, but are
not in contact. However, under the action of $\delF$ these beads do
come into contact. The experimental results indicate that this is not
an important effect in understanding the response of the system. These
contacts produce a force that only acts during part of the
oscillation cycle. If this effect were important, then the response of
the accelerometer would not be sinusoidal, but would take on some more
complicated wave form. Thus, in order to understand the experimental
results described above, I will not consider these ``sometimes'' bonds.

Finally, one might object to the use of a Eq.\ \hertz\ on the basis
that it ignores the effect of the surface roughness of the beads.
However, the essential idea that for small oscillations the contact
will behave like a spring does not depend on the fact that the contact
be of the Hertzian type. All that is required is an analytic form for
the behavior near equilibrium in order to do an expansion for small
$\delh$. The dependence then of $\eqK$ on the equilibrium configuration
may then become more complicated; however, I will not need the analytic
form for $\eqK$ in this study.

In terms of the response to small oscillations, the system can then be
viewed as a collection of balls connected by springs, with the spring
constants determined in some way by the force network of the
equilibrium system. When the amplitude of a ball changes as a function
of time, this indicates that the equilibrium configuration of the
system has changed -- the network has relaxed. The experimental
results then raise some very interesting questions: What information
about the force network is reflected in \figs\exptime\ and
\xfig\expfreq? How does the network relax in time? What is the role of
the geometrical disorder in the system? These are the questions that I
address in this study.

I focus on an an extremely simple model for the system. I consider a
network embedded in two spatial dimensions. I ignore the positional
disorder, and consider the network topology to be that of a square
lattice. I choose a velocity dependent dissipation with a uniform
damping constant. The only disorder present in the system is in the
distribution of spring constants, $\eqK$. This paper will compare the
response of this model system to the response observed in the
experiments.

I now present the mathematical formulation of this ball and spring
model.  Consider the beads, confined to a two dimensional square
lattice as in \fig\sprlat{Basic ball and spring model for the system.
The wall on the right oscillates to provide the driving force.}. I let
the driving force be a sinusoidal motion of one entire wall.  Because
energy is constantly being pumped into this system by the vibration,
there must be some method for dissipation in the system. I assume a
damping force that is linearly dependent on the velocity difference
between two adjacent beads. The final simplification comes from an
assumption of the decoupling of the motion in the $x$ and $y$
directions. This will be true in the limit of very small oscillations.

In this model, the displacement $\up$ of a bead at position $p = (x,y)$
is given by the differential equation
\eqn\timeeq{
m{d^2\up \over dt^2} = \sum_{\pp} \eqK(p,\pp) (\upp - \up) +
\sum_{\pp} \beta \left({d \upp \over dt} - {d \up \over dt}\right),
}

\ni where $m$ is the mass of a ball, $\pp$ is a nearest
neighbors to point $p$, $\beta$ is the damping constant, and $\eqK(p, \pp)
= \eqK(\pp, p)$ is the spring constant between $p$ and $\pp$.

For boundary conditions, I take that that $\up = 0$ on all walls that
do not supply a driving force. For the oscillating wall, I have the
form $u = \exp(i\omega t)$, where, without loss of generality, I have
taken the amplitude of the wall's vibration to be unity.

It is possible to take Eq.\ \timeeq\ as the starting point for
numerical investigation by merely integrating this equation to find
out the response of each ball as a function of frequency. However,
this does not take advantage of the fact that for small amplitude
vibrations, the harmonic driving force produces a harmonic motion of
the particles in the bead pack.  Thus, I know that the solution of
interest takes the form $\up = \Ap
\exp(i\omega t)$, where $\Ap$ is the complex amplitude for the
vibration at point $p$.  Combining this with Eq.\
\timeeq, I find
\eqn\ampeq{
m\omega^2\Ap + \sum_{\pp} (\eqK(p, \pp) + i\omega \beta) \left(\App -
\Ap\right) = 0.
}

\ni The boundary conditions for this equation is that $\Ap = 0$
on the three fixed boundaries, and that $\Ap = 1.0$ on the oscillating
wall. This equation described the steady state response of this
ball and spring model. To find the frequency response for a given
system, it is necessary to know the values of $\eqK(p, \pp)$.

\newsec{The Continuum System}

If I consider the case of infinitesimally small particles (i.e.\ the
continuum limit), it is possible to rewrite Eq.\ \ampeq\ in terms of a
differential equation.  The resulting equation can be solved exactly
for the case of a uniform spring constant.  Letting $\eqK(p, \pp) =
\Kz$ for all $p$ and $\pp$, I write $\Ap$ as $\Axy$ and consider a box
with $x$ and $y$ dimensions $L_x$ and width $L_y$, respectively. For
such a system, \ampeq\ becomes
\eqn\diffeq{
m\omega^2\Axy + (\Kz + i\omega\beta)\nabla^2\Axy = 0.
}

\ni The boundary conditions for this equation are  $A(L_x, y) = 1.0$,
and $A(x, 0) = A(x, L_y) = A(0, y) = 0.0$. The solution to this
equation, with these boundary conditions can be obtained by standard
separation of variable techniques, and I find
\eqn\axysol{
\Axy = {4 \over \pi}\sum_{n=0}^{\infty} {\sin[(2n + 1)\pi y/L_y]\over
(2n + 1)} \left({\sinh(\aln x / L_x) \over \sinh(\aln)}\right),
}

\ni where $\aln$ is defined by
\eqn\alndef{
\aln^2 / L_x^2 = (2n + 1)^2 \pi^2/ L_y^2 - {m\omega^2\over\Kz +
i\omega\beta}.
}

It is very simple to sum this series numerically, and find the
amplitude as a function of frequency for any point in the box. All
that is necessary is to choose values for the parameters $m$, $\Kz$,
$\beta$, $L_x$ and $L_y$. In the solutions shown here, I choose $m =
1$, $\Kz = 1$, $L_x = L_y = 1$. \nfig\exactpl{Frequency response of a
the homogeneous system: (a) shows the response of the system
with a damping coefficient $\beta = 0$, (b) shows the response
with damping coefficient $\beta = .003$.} Figure \xfig\exactpl\ shows
the phase and frequency response at $\xy = (.47, .52)$ for two
different values of $\beta$. In \exactpl(a), I show the results for
$\beta = 0$.  There is a set of frequencies where the amplitude at
$\xy$ is peaked, and is much greater than the amplitude of the driving
wall.  Additionally, there are frequencies at which the amplitude
jumps discontinuously. The frequencies, $\ompk$, at which these peaks
and discontinuities occur are well described by the expression
\eqn\simpfreq{
\ompk^2 = k\pi^2(n_1^2 + (2n_2+1)^2),
}

\ni where $n_1 > 0$ and $n_2 > 0$ are both integers. The frequencies
defined by \simpfreq\ are shown in \exactpl(a) by the pluses located
on the $x$-axis.

The importance of these frequencies can be understood in two ways. The
first comes from a direct examination of \axysol\ and \alndef. At
$\omega = \ompk$, the value of $\alpha_{n_2} = n_1\pi i$. For this
value of $\aln$, the function $\sinh(\aln)$ vanishes, and the
contribution for that the $n=n_2$ term in the series \axysol\ goes to
infinity.

However, the frequencies of \simpfreq\ also correspond to natural
frequencies of the unforced system. Let $\beta = 0$ and $A(L_x, 1) =
0$. In this case, Eq.\ \diffeq\ becomes an eigenvalue problem, and the
eigenfunctions, $\eigfun$, are
\eqn\eigenfunc{
\eigfun = \sin(2q_1\pi x/L_x)\sin(2q_2\pi y/L_y),
}

\ni with the corresponding eigenvalues, $\omega_i^2$, given by
\eqn\eigvalues{
\omega_i^2 = k\pi^2(q_1^2 + q_2^2),
}

\ni where again $q_1>0$ and $q_2>0$ are both integers. Not all
eigenfrequencies given by Eq.\ \eigvalues\ are manifested in the
system. This is due to the symmetry of the driving force -- it is
symmetric about a horizontal axis at the center of the box.  This
excludes all of the eigenmodes which are asymmetric about this axis
(i.e.\ have an even value for $q_2$).

Figure \xfig\exactpl(b) shows the response for a finite value of
the damping coefficient, $\beta = .003$. The frequencies defined by
\simpfreq\ are again shown as pluses along the $x$-axis. There are two
distinct differences between this response curve and \exactpl(a). The
first is that there is an overall decrease in response as the
frequency increases. This reflects the fact that the damping force is
proportional to $\omega$, and thus the damping force gets larger at
higher frequencies.  By looking at a larger range in frequency, it is
clear that this trend is nothing as straight-forward as a simple
exponential decay.

The second interesting feature is the set of peaks and valleys which
are superimposed on this decreasing response curve. The lowest
frequency excitations are still identifiable as distinct entities with
peaks clearly occurring at the resonance frequencies.  The higher
frequency modes, on the other hand, can not be distinguished in the
response curve. The finite value of the damping term causes the
resonance peaks to be greatly decreased in magnitude, and to be much
broader in terms of their frequency response. At higher frequencies,
many broad peaks are superposed to get the measured response of the
system. At some frequencies the result is a slightly higher than
average response, and at other frequencies, it is slightly lower. This
causes the apparent small peaks and valleys at high frequencies. At
even higher frequencies, the effects of the damping become dominant,
and the curve becomes smooth.

This is indeed reminiscent of the frequency response in the
experimental system. There is a general trend of a decreasing
amplitude with increasing frequency, with peaks and valleys
superimposed on this decreasing curve.  In the experiments, these
oscillations are much more striking than in this uniform system, but
the behavior is the same.

\newsec{The discreet model}

I now study Eq.\ \ampeq\ numerically, looking at the properties of the
solution for a particular distribution for the values of $\eqKpp$.
The method that I use to find the solutions of equation \ampeq\ is the
Bi-Conjugate Gradient Method \bcgref. This method determines solutions
to a matrix equation of the form
\eqn\bcgeq {
Mx = b,
}

\ni by finding the minimum of the function
$g(x) = {1\over2} \overline{x} \cdot M \cdot x + b \cdot x$. In order to
transform Eq.\ \ampeq\ into the form of \bcgeq, it is necessary to
write \ampeq\ as an equation for its real and imaginary parts. As a
result, the matrix $M$ is not symmetric, and this is why the
bi-conjugate, rather than the conjugate, gradient method must be used.
In my simulations, the amplitude for each point is accurate to 1 part
in 10000.

It is, of course, possible to simply invert the matrix A to find the
solution. However, this does not take advantage of the sparseness of
the matrix. All of the calculations for the results shown in this
paper were done using a Sun SPARCstation {\sl IPX} using approximately
150 hours of CPU time.

All that remains is too choose a distribution for the set of
$\{\eqKpp\}$. For simplicity, I take a random distribution of values
on the interval $[0, K_{max}]$.  The results that I show are for a
square array of balls of size $20 \times 20$.
\nfig\sprpic{The $20\times 20$ ball and spring system used for these
simulations. The intensity indicates the strength of the bond
connecting two balls: black indicates that $\eqK = \Kmx$, white
indicates $\eqK = 0$, with a linear gray scale for intermediate
values. The large black ball at (8, 10) indicates the location where
all of the measurements occur.} Figure \xfig\sprpic\ shows the spring
configuration for the array that has been used for the results
presented below.  Black indicates that $K_h = K_{max}$, while white
means $K_h = 0$, with a linear grayscale for intermediate values.

I choose $m = 1$, $\Kmx = .25$, $\beta = .003$ and measure the
response in the frequency range from $\omega = (.01, 1.3)$. I shall
show below how to assign physical units to these values.
\nfig\freqres{Frequency response of ball at (8, 10) as a function of
frequency: (a) for low frequencies, (b) for intermediate frequencies
and (c) for high frequencies. The pluses along the $x$-axis indicate the
frequencies of the normal modes.} Figure \xfig\freqres\ shows the amplitude
for a ball near the center of the pack of the springs (the precise
position of the ball is indicated in \sprpic).

There are three distinct regimes of behavior. At very low frequencies
(\freqres(a)), the behavior is very similar to the response seen at low
frequencies in the continuum case. There are frequencies at which
peaks occur, and these peaks show little overlap.

In the intermediate frequency range (\freqres(b)), the response of the
system is qualitatively the same as that shown in the experimental
results of
\expfreq. There is an overall decrease of the response, and an
irregular set of peaks and valleys.

For $\omega > 1$ (\freqres(c)), the response becomes very smooth and
decreases very quickly. The change in response at $\omega \approx 1$ can be
understood intuitively.  Consider a single ball attached to four
springs all with $\eqK = \Kmx$. Such a ball will have its fundamental
excitation at a frequency
\eqn\omxdef{
\omx = \sqrt{4 \Kmx \over m}.
}

\ni This then is the ``natural" frequency (known as the Einstein
Frequency \phonref) for a ball which, by chance, happens to be
surrounded by springs all of which have $\eqK \approx \Kmx$. Equation
\omxdef\ then defines the highest frequency at which any bead will
naturally oscillate. Any higher frequency oscillations would require
every neighbor to vibrate in opposition with many of its neighboring
particles. Below it is shown that this results in a distinctly
different character to the system's oscillations.

For the values of $\Kmx$ and $m$ given above, $\omx = 1.0$. For other
simulations, with different values for $\Kmx$, I found that this
change in the frequency response of the system occurred always at the
frequency given by Eq.\ \omxdef. This frequency determines the
frequency scale for the system, and I will consider my frequency as
measured in units of $\omx$. Experimentally, $\omx$ might be measured
by finding the frequency at which this rapid decrease in the response
occurs. It is also possible to estimate a lower limit for $\omx$ in
the experimental system by assuming the Hertzian contact law and that
contact force between any two beads is not significantly greater than
the weight of a single bead. Making use of some of the information
about the beads given in ref.\ \xref\prlliu, and Eqs. \hertz,
\basespr\ and \omxdef, I find that
\eqn\omxest{
\omx > 45 {\rm kHz},
}

\ni which is above the observed range in the experiment.

Consider now the phase of a given ball in the array.
\nfig\phaseres{Phase of a single ball for all driving frequencies.
There is a linear regime in the intermediate frequency range. The
pluses along the $x$-axis mark the frequencies of the normal modes.}
Figure \xfig\phaseres\ is a plot of the phase versus frequency for the
same ball as was used in \freqres.  Again there is distinctly
different behavior for $\omega > \omx$. There is also a linear regime
in the phase for $.3\omx < \omega < \omx$. A linear relationship
between the observed phase and the driving frequency is seen
experimentally \liuunp. A simple calculation shows that for a constant
phase velocity, $\vph$, the relationship between the frequency of the
oscillation, and the observed phase at a point should be linear. If
the slope of the phase response is $\slope$, then the phase velocity
is given by
\eqn\phsvel{
 \vph = X / \slope,
}

\ni where $X$ is the distance from the source of the driving to the
point of observation. In the data shown here, the value of $X$ is 12
particle diameters. In the experimental system, the particle diameter
is $.5 {\rm cm}$. Using this value and the slope from \phaseres, I
find that
\eqn\velobs{
\vph \approx .03 \omx,
}

\ni where $\omx$ is measured in ${\rm sec}^{-1}$, and the
velocity is given in units of cm/sec. In the experimental system,
$\vph = 57 {\rm m/sec}$. Matching to that velocity implies
that
\eqn\expomx{
\omx \approx 200,000 {\rm sec}^{-1},
}

\ni which is clearly beyond the experimental regime. With this value
of $\omx$, \freqres\ indicates that the typical $\Delta \omega$ over
which the amplitude changes appreciably is $\approx 20$ kHz. This
compares with the experimental value which is quoted as $\Delta
\omega \approx 3$ kHz.

The parameters $m$ and $\Kmx$ determine the value of $\omx$ through
\omxdef. The only parameter whose effect I have not discussed is the
damping constant $\beta$.  The value observed for $\slope$ depends
sensitively on the value chosen for $\beta$.  If a smaller value for
$\beta$ is chosen, the value of $\slope$ is increased. Additionally, a
smaller $\beta$ results in many more small peaks in the frequency
response data, but the sharp decreases are much shallower. If, on the
other hand, the value of $\beta$ is increased, the sharp decreases in
the response are much deeper, but they occur with a much larger
spacing in frequency.  The effects on the peaks can be understood my
considering the results from the normal mode analysis below.

Because the normal modes proved to be so important in the response of
the continuum system, I consider their manifestation in the discreet
case. I again use the configuration of \sprpic.  The normal modes are
calculated for the undamped system with the zero amplitude boundary
conditions at all walls. The numerical calculation of the eigenmodes
is done using the appropriate subroutines from LINPAK.
\nfig\modepic{Typical modes for the system in three frequency ranges:
(a) $\omega = 0.1013\omx$,; the mode is extended; (b) $\omega =
.6053\omx$; the mode shows some localization; (c) $\omega =
1.000\omx$; the mode is very localized.} Figure \xfig\modepic\
shows three eigenmodes for the system. The first one is the second
lowest frequency mode of the system, and shows little trace of the
local non-uniformities of the network.  The second mode is in the
middle of the spectrum at $\omega = .6053\omx$.  The picture shows
some localization of the oscillation, but not complete. The final mode
shown has $\omega = 1.000 \omx$, and there is striking localization of
this mode. This is true of all the modes at high frequency, and this
is the reason for the significant change in the response function at
driving frequencies with $\omega > \omx$.

These normal mode frequencies are shown as pluses in
\figs\freqres\ thru \xfig\phaseres. Notice, as in the continuum case,
that at low frequencies the sparseness of the normal mode allows them
to be observed as distinct excitations of the system. At higher
frequencies, however, the modes overlap and there can be no
one-to-one identification between the structures in the response and a
given normal mode.

Note also that the density of modes becomes very low for frequencies
greater than $\omega \approx \omx$. From the results for
low frequencies, this would seem to suggest that these modes should
be seen distinctly in the response curve. However, because these modes
are highly localized, they will only be seen in the response if
the detector happens to be inside the region of space where the mode
has a non-zero value. In addition, these highly localized modes must
have some overlap with the boundary or they cannot be significantly
excited by the driving force.

The major difference in this non-uniform system is that not all modes
will contribute equally to the system's response. A high frequency
mode which has its amplitude localized far from the driving wall
is much less important than a mode which is localized near
the driving wall. Thus, not only are the functional forms of the
normal modes more complicated, it is also important to know which
modes make the dominant contributions at each frequency.

In order to measure the contribution of each mode to the system
response, I use the set of normal modes as a set of basis vectors to
describe the amplitudes of the balls at each driving frequency. At a
given driving frequency $\omega$, the amplitude of the oscillations is
given by $\Axy$. $\Axy$ can be written as a superposition of the
eigenmodes,
\eqn\Axydec{
\Axy = \sum_{i = 0}^{N\times N} c_i \fixy,
}

\ni where $c_i$ is given by
\eqn\cidef{
c_i = \sum_{j = 0}^N\sum_{k=0}^N A(j,k)f_i(j,k).
}

\ni Thus, by plotting the $|c_i|$ as a function of frequency,
I can see when each mode is contributing to the behavior of the driven
system.

With the modes calculated numerically, it is a simple matter to
calculate the values of $c_i$ for each driving frequency.
\nfig\onefreq{Values for the $|c_i|$'s as a function of the mode
frequency $\omega_i$. The driving frequency is $\omega = .6\omx$.} Figure
\xfig\onefreq\ shows the values of the $|c_i|$ versus $\omi$ at a
driving frequency $\omega = .6 \omx$. While it is clear that the most
important contributions come from the modes that have $\omega_i
\approx \omega$, the structure of the curve is
complicated. It is also clear that the level of a modes excitation is not
simply determined by the value of $|\omega - \omega_i|$.

There are, however, some general patterns to these $c_i$'s.
\nfig\maxci{The most important modes. This figure shows the value of $\omega_i$
versus the driving frequency $\omega$ for the mode which
has the largest value of $|c_i|$ at each value of value for $\omega$.
Note that above $\omega = \omx$, the linear behavior starts to break
down and the curve becomes very irregular.} Figure
\xfig\maxci\ shows the frequency of the mode which has the largest
value of $|c_i|$ at each driving frequency.  This is a linear function
with slope 1, and indicates that the most important mode is one which
has its resonant frequency near the driving frequency. Again, it is
clear that the behavior sharply changes at $\omega \approx \omx$.

Finally, I consider the $c_i$'s for several eigenmodes which have
their frequencies very close together. Shown in \fig\cifreq{The value
of $|c_i|$ as a function of the driving frequency $\omega$ for three
successive modes of the system.} are the values of the $|c_i|$ as a
function of frequency for three modes with a resonance frequency near
$\omega = .6\omx$.  Each curve shows a clear resonance behavior when the
driving frequency approaches its natural frequency. However, it is
clear that each mode is excited to a different degree by the driving
force.

As was stated above, it is possible to understand the changes in the
response curve as the value $\beta$ is changed by considering the
normal mode.  When the value of $\beta$ is increased, each of these
modes will have a much sharper peak in its $c_i$ when the driving frequency
approaches the mode's natural frequency. Thus, any mode which has a
non-zero contribution to the amplitude at the detector is more likely
to make its presence felt. Thus, there are many more smaller peaks in
the measured response.

\newsec{Toy Model for Relaxation}

Because \ampeq\ is a steady state solution for fixed $\omega$,
there will be only one amplitude measured for all time. I have allowed
for no relaxation of the lattice which is the source of the behavior
in \exptime. As a toy model for this relaxation process, I consider
the amplitude at one point, while randomly changing one bond per unit
of time, $\tau$. This bond is assigned a new random value between 0 and
$\Kmx$.  Having changed one of the values of $\eqK$, I then find the
new solution of \ampeq. This new solution has a different value
for the observed amplitude, and by plotting this as a function of
elapsed time, I generate a time sequence. I then compare the power
spectra of this series to the experimental power spectra.

I begin the simulations with the same bonds configuration as shown in
\sprpic, and measure the amplitude of the same ball as I used for the
frequency data. The system is driven with an oscillation of frequency
$\omega = .6 \omx$, with a value of $\beta = .01$.
\nfig\timeres{Results for the time trace of the amplitude as the bond
network is changed: (a) is a sample of the time trace data, (b) shows
the corresponding power spectra. A line with slope 2 is shown for
comparison.} The amplitude of the vibration as a function of time is
shown in
\timeres(a). Using a much longer time series, I can calculate the
power spectrum, $S(f)$, for this time series \numrec, with the results
shown in figure
\timeres(b). There is a clear power law behavior in the power
spectrum. If the power spectrum is written as $S(f) \sim
f^{-\powspc}$, then I find that $\psi \approx 1.9$ over a region of
one and a half decades. This is fair agreement with the experiments,
which find values for $\psi$ between 2 and 2.2.

It is not surprising that $\psi \approx 2$ in both the toy model and
the real system for the following reason \prlliu. Imagine that at one
time the amplitude of the detector has some value $B_0$ and then one
of the bonds is changed. This will either raise or lower the observed
amplitude by $\Delta B_0$, and I might expect, in general, that the
change might be equally likely to cause an increase as a decrease.
Then another bond is changed, and the amplitude changes by some amount
$\pm \Delta B_1$. In this picture it is clear that the amplitude is
executing a random walk. Such a time trace is known to have a power
spectrum with an exponent $\powspc = 2$.  Thus, the power law behavior
is not surprising, and the exponent near 2 seems to be
intuitive.

There are, however, several assumptions in this argument which are
unjustified. Are $B$ and $\Delta B$ actually uncorrelated? Are
successive values for $\Delta B$ uncorrelated? In order to understand
the experimentally observed results, a more physically motivated
process for the network relaxation is necessary.

\newsec{Conclusions and Discussion}

The principal objective of this paper is to show that the vibrational
properties of a granular material can be reproduced with a simple
linear ball and spring model.  I have been able to reproduce
qualitatively the frequency response of a single grain to a sinusoidal
driving frequency.  There is as yet no way to quantitatively compare
with experiments, because there is no well-defined method for
characterizing the statistical properties of the ``noisy'' response
curve. The normal mode analysis indicates that the peaks in the
spectrum cannot be associated on a one to one basis with a resonance
of the system; however, it is clear that the behavior at a given
frequency is a results of a superposition of the normal modes which
have their resonance frequencies in the regime of the driving
frequency. The toy model for the relaxation of the lattice does
reproduce a power law in the behavior in the power spectrum. The fact
that the model is simple, and yet produces the experimentally observed
behavior, suggests that it may be worth studying in its own right.

However, there is at least one aspects of the real system that can not
be studied with this simple ball and spring model -- the experiments
shows an extreme sensitivity to thermal fluctuations. In additional
measurements, Liu and Nagel place within the sample a small heater the
size of a single bead, and then run a heat pulse through it \liuunp.
The pulse changes the temperature of single bead by approximately 0.8
K, and they find that the response of the distant accelerometer
changes almost instantaneously by 25\%. The effect of the change in
temperature is to change, by thermal expansion, the size of the beads
in a region near the heater. The change in radius of the bead is
estimated at 1000 \AA.  In order to understand these temperature
effects, a much more physically motivated model is needed for the
granular system.

The major defect with the current model is that it does not allow for
a general topology for the force network. Also, the spring constants
are clearly determined by the different equilibrium forces on the
beads. However, there is no requirement that these forces sum to zero,
as they should in equilibrium. This problem can only be solved by
choosing a specific form for the force law between touching spheres.
Once these two problems are solved, it would be straight forward to
study the effects of local thermal fluctuations.

\bigskip
\goodbreak
\leftline{\bf Acknowledgments}

I would like to gratefully acknowledge Thomas Witten and Jysoo Lee for
invaluable conversations and suggestions. I would also like to thank
Chu-heng Liu and Sid Nagel for providing me with the latest results
from their experiments, and for many useful discussions. I also thank
T.\ C.\ Halsey, M.\ Marder, H.\ Herrmann, M.\ Gannon, C.\ Moukarzel,
K.\ Lauritsen and S.\ Melin.

\vfil
\eject
\headline{\hfil}

\listrefs
\listfigs
\bye